\documentclass[fdp,a4paper,fleqn
]{w-art}
\usepackage{times,cite,w-thm}
\theoremstyle{plain}

\theoremstyle{definition}

\usepackage[]{graphicx}

\def\be{\begin{equation}}
\def\ee{\end{equation}}
\def\bea{\begin{eqnarray}}
\def\eea{\end{eqnarray}}

\newcommand\fverb{\setbox\pippobox=\hbox\bgroup\verb}
\newcommand\fverbdo{\egroup\medskip\noindent%
                        \fbox{\unhbox\pippobox}\ }
\newcommand\fverbit{\egroup\item[\fbox{\unhbox\pippobox}]}

\newcommand{\bear}{\begin{eqnarray}}

\newcommand{\eear}{\end{eqnarray}}

\newcommand{\bsea}{\begin{subeqnarray}}
\newcommand{\esea}{\end{subeqnarray}}
\newbox\pippobox

\newcommand{\ga}{\gamma}

\newcommand{\da}{\delta}

\def\6{\partial}

\def\a{\alpha}
\def\g{\gamma}

\def\m{\mu}
\def\n{\nu}

\def\s{\sigma}

\def\sq
\def\a{\alpha}

\def\d{\delta}

\newcommand{\Figref}[1]{Fig.\ref{#1}}

\begin{document}
\DOIsuffix{theDOIsuffix}


\title[Dilatonic Transport]{Charged Dilatonic Black Holes and their Transport Properties}

\author[B. Gout\'{e}raux]{Blaise Gout\'{e}raux\inst{1,}}
\address[\inst{1}]{APC, Univ. Paris-Diderot, CNRS UMR 7164, F-75205 Paris Cedex 13, France}

\author[B. S. Kim]{Bom Soo Kim\inst{2,3,}
\footnote{Corresponding author\quad E-mail:~\textsf{bskim@physics.uoc.gr},
            Phone: +30\,2810\,394\,265,
            Fax: +30\,2810\,394\,274}}
\address[\inst{2}]{IESL-FORTH, 71110 Heraklion, Greece}

\author[R. Meyer]{Ren\'{e} Meyer\inst{3,}}
\address[\inst{3}]{\mbox{Crete Center for Theoretical Physics,
Department of Physics, University of Crete, 71003 Heraklion, Greece}}

\begin{abstract}
We briefly explain the consistency conditions imposed on the effective holographic theories, which 
are parameterized by two real exponents $(\gamma,\delta)$ that control the IR dynamics.
The general scaling of DC resistivity with temperature at low temperature 
and AC conductivity at low frequency across the whole $(\gamma,\delta)$ plane are explained. 
There is a codimension-one region where the DC resistivity is linear in the temperature.  
For massive carriers, it is shown that when the scalar operator is not the dilaton, 
the DC resistivity  scales as the heat capacity (and entropy) for $(2+1)$-dimensional systems.
Regions are identified where the theory at finite density is a Mott-like insulator. 
This contribution is based on [C.~Charmousis, {\it et al.} JHEP {\bf 1011}, 151 (2010)] 
\cite{Charmousis:2010zz} with emphasis on the transport properties of charged dilatonic black holes
with potential.
\end{abstract}
\maketitle

\section{Introduction}

In recent years, AdS/CFT correspondence opens up new ways to investigate low temperature dynamics of 
strongly correlated electron systems \cite{s1}. The physics of condensed matter system (CM) is 
very diverse, thus we need to find various theoretical models and to classify them in terms of its 
universality classes at low temperature. Classification criteria may include the appropriate phases 
via their thermodynamic properties including phase transitions  
and their transport coefficients, which is directly connected to what is measured by experiments. 

One of the first and comprehensively investigated example of such classes is charged Reissner - Nordstr\"om 
(RN) black holes of Einstein-Maxwell gravity with a cosmological constant. Such solutions proved to be 
the simplest laboratory for the study of phase transitions at finite density \cite{RN1}, fermionic 
quasi-particles with non-Fermi liquid behavior \cite{za}, as well as an emerging scaling symmetry 
at zero temperature \cite{mc2}, and superfluid (superconductivity) phase transitions \cite{super1}. 
While this model acquired interesting and important developments, it has a large entropy at 
extremailty signaling instability. Anticipating CM application, it is desirable to find many more models,
which have qualitatively distinct features with each other.     

A family of charged black hole solutions with one parameter to the low energy string theory with dilaton 
has interesting and several distinctive thermodynamic properties \cite{Gibbons:1987ps}: 
The Maxwell field ($F$) in the solution has a generalized coupling with the dilaton ($\phi$) as 
$e^{-2 \alpha \phi} F^2 $, where $\alpha$ is a parameter. When $\alpha = 0$, it is nothing but the 
RN black hole, which has vanishing extremal temperature $T=0$ and finite entropy $S=S_0$ at extremality. 
The extremal behavior of the solution changes according to the parameter $\alpha$ : 
The extremal entropy vanishes for non-zero $\alpha$, while the extremal temperature changes as 
$T=0$ for $0 < \alpha <1$, $T=T_0$ for $\alpha = 1$ and  $T=\infty$ for $\alpha>1$.    
Thus it is natural to expect that we can generate several qualitatively distinct classes with a dilaton potential and 
a coupling constant to the Maxwell field with a scalar dilation. 

\section{Effective Holographic Approach}

We would like to model strongly-coupled holographic systems at finite charge density 
by including a leading relevant scalar operator, which is generically uncharged 
and drives the renormalization group flow of the system from UV to IR. 
The corresponding holographic theory is an Einstein-Maxwell-Dilaton system with a scalar potential. 
In the case of zero charge density such theories have been analyzed (see {\it e.g.} \cite{Gursoy:2007cb})
and generalized to the finite charge density case (see {\it e.g.} \cite{Gubser:2009qt}).
This is a phenomenological approach based on the concept of Effective Holographic Theory (EHT), 
in analogy with Effective Field Theories which is low energy approximations to QFT.
It allows a parametrization of large classes of IR dynamics and a survey of important observables, 
while it is not obvious if concrete EHTs can be embedded in well-defined string theories, or 
if classical singular solutions are acceptable as saddle points of the dual CFTs 
\cite{Gubser:2000nd,Gursoy:2007cb,Heemskerk:2009pn}.

Our strategy is to select a minimal set of operators (or dual string fields) that are expected to dominate 
the low energy dynamics and then parametrize their EHT in terms of a general two-derivative action. 
This set is composed of the bulk metric, $g_{\m\n}$, controlling the energy distribution, bulk gauge field, $A_{\mu}$, 
associated with a conserved U(1) current $J_{\mu}$ and a scalar leading-relevant operator, $\phi$, controlling 
the coupling constant and its running in the IR.
For a CM system, $\phi$ may represent the strong interactions of the ion lattice  
or the effect of spins on the charge carriers. If the charge carriers are dilute, we expect that 
the dynamics can be treated in the probe approximation, while the full set of equations must be solved 
at sufficiently large doping.

At the two-derivative level, the EHT in the Maxwell case is described in the Einstein frame by the action
\be
S= M^{p-1}\int d^{p+1}x\sqrt{-g}\left[R-{1\over 2}(\partial\phi)^2+V(\phi)\right]
-M^{p-1}\int d^{p+1}x\sqrt{-g}{Z(\phi)\over 4}F_{\m\n}F^{\m\n}
\label{ActionLiouville}\ee
Indeed, the most general, $p+1$ dimensional, two-derivative action containing a metric, a vector and a single scalar can be brought to this form via field redefinitions. 
On the other hand, the linear Maxwell action, $S_{Max}$, must be replaced by the DBI action 
$S_{DBI}= -M^{p-1}\int d^{p+1}x e^{2k\phi} Z(\phi)\left(\sqrt{-\det(g+e^{-k\phi}F)}\right. $ $\left.-\sqrt{-\det g}\right),$
in the fundamental charge case.
Here $g = e^{-k \phi} g^\sigma$ is the Einstein frame metric. $Z$ has been normalized such that 
the DBI action reduces to the Maxwell one for $F^2\ll 1$. The parameter $k$ encodes the dilatonic 
nature of the scalar field $\phi$. If $\phi$ is the dilaton, $k={-\sqrt{2\over p-1}}$. 
Even though the discussions can be generalized to other dimensions straightforwardly, we would like to 
concentrate on the 4 dimensional case, targeting (2+1)-dimensional CM phenomena. 

While, in holographic theories, it is standard to assume that there is a non-trivial fixed point in the UV to ensure 
a globally controllable behavior with an appropriate UV boundary, this assumption is not necessary if our only interest 
is to understand the infrared (IR) dynamics.\footnote{In the holographic approach, This direction is explicitly pursued recently 
in \cite{Bredberg:2010ky}} For example, the non-trivial dynamical condition, 
which defines the ``vacuum" and expectation values, is imposed in the IR typically by a regularity condition 
in the far IR. For this purpose and also motivated by controlled examples, 
we model $V(\phi)=V_0 e^{-\delta\phi}$ as generic scalar operators have non-trivial n-point functions in all realistic examples.
We further demand that $e^{-\delta\phi}\to 0$ in the UV to guarantee that simple 
modifications of the potential like $V\to {p(p-1)\over \ell^2}+V_0 e^{-\delta\phi}$ will lead to asymptotically AdS solutions without 
modification of the IR behavior. We also model $Z(\phi)\sim e^{\gamma\phi}$ motivated from both gauge supergravity actions stemming 
from string theory as shown in the introduction. 

\begin{figure}
\sidecaption
\includegraphics[width=45mm,height=40mm]{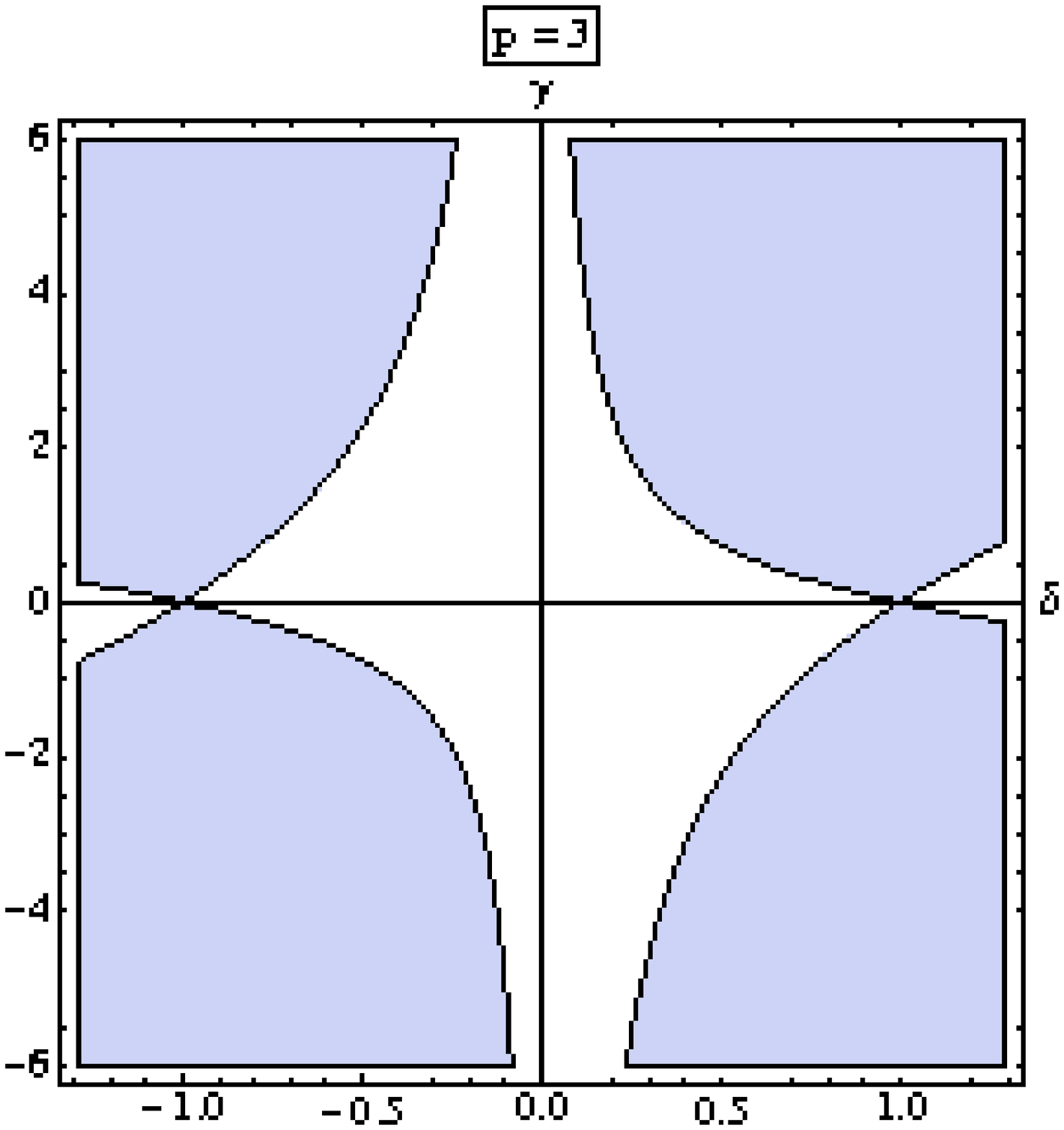}
\includegraphics[width=45mm,height=40mm]{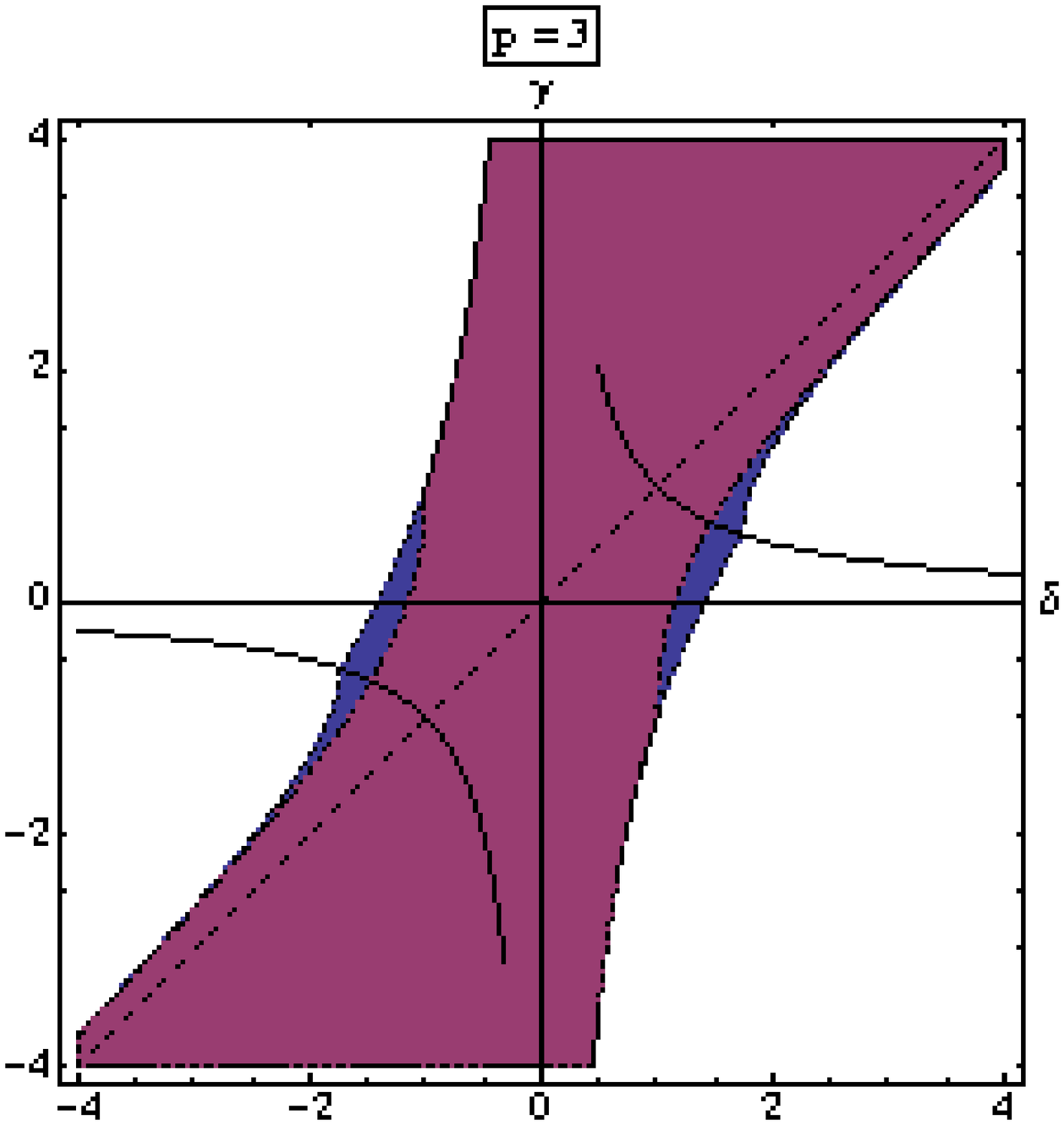}%
\caption{$\bullet$ Uncharged case (left): Allowed regions of the parameter space from spin 2 and spin 1 fluctuation analysis.  
$\bullet$ Charged case (right) : The solid black and dotted black lines are $\gamma\delta =1$ and $\g=\d$, respectively. 
The outer blue region is allowed region by Gubser's criteria and the inner purple region is that after taking care of all 
the constraints including the allowed spectrum and thermodynamic stability.}
\label{Constraint}
\end{figure}

With this EHT setup, we generically encounter naked singularities deep in IR. 
Thus we investigate the consistency condition of our EHT model in IR with the following listed criteria and constrained 
the parameter space $(\gamma, \delta)$, which is depicted in \Figref{Constraint} :  
\begin{itemize}
\item Gubser's Criteria : {\em A naked singularity is unphysical if the scalar potential is not bounded 
below when evaluated in the solution}\cite{Gubser:2000nd}. In holography, well-defined and controllable perturbations of
N=4 sYM lead to mild naked singularities in the bulk, which are generically resolved by lifting to higher dimensions 
or eventually by the inclusion of the stringy states. Thus not all naked singularities are unacceptable. 

\item Well-defined spectrum : {\em A naked singularity describes unreliable holographic physics
 if the second-order equations describing the spectrum of small fluctuations around the solution are
  not well-defined Sturm-Liouville problems}\cite{Gursoy:2007cb}. We change our low energy fluctuation equations into 
  Schr\"odinger form and require our solutions to have only one normalizable mode in the deep IR  
  for spin 1 and spin 2 spectra. This also allows us to classify the nature of low energy spectrum, which is 
  discussed when we discuss the transport properties. 

\item Thermodynamic Stability : We accept the AdS completable solutions in the thermodynamically unstable case as well as 
  thermodynamically stable solutions. Please consult the details of the thermodynamic properties of the various solutions 
  in the companion contribution \cite{Rene} of these proceedings.    
\end{itemize} 

\vspace{-0.1in}
\section{Transport Properties}

Conductivities can be calculated by linear response theory using a Kubo formula from the two-point function 
of the fluctuations of the gauge field. We will consider long wavelengths compared to their frequency, which 
is a good approximation for optical CM experimental measurements.
Following \cite{Horowitz:2009ij,Goldstein:2009cv}, we calculate the zero-temperature asymptotics of the AC conductivity 
$\sigma(\omega)\sim \omega^{2|\nu|-1}$ by 
transforming our fluctuation equations into a Sch\"odinger equation with potential $V(z)={\nu^2-{1\over 4}\over z^2}$ 
in the near horizon region. 
While the DC conductivity is the zero frequency limit of the AC conductivity, this limit may be ill-defined \cite{Romatschke:2009ng}. 
A direct calculation can be done by turning on a constant electric field, calculating the induced current 
and thus obtaining the ratio \cite{Karch:2007pd} using the probe DBI action which describes massless, self-interacting 
but non-backreacting charge carriers.
We also summarize the low energy spectrum of our models, which was analyzed fully in \cite{Charmousis:2010zz}.

\begin{figure}
\begin{minipage}{35mm}
\includegraphics[width=\linewidth,height=30mm]{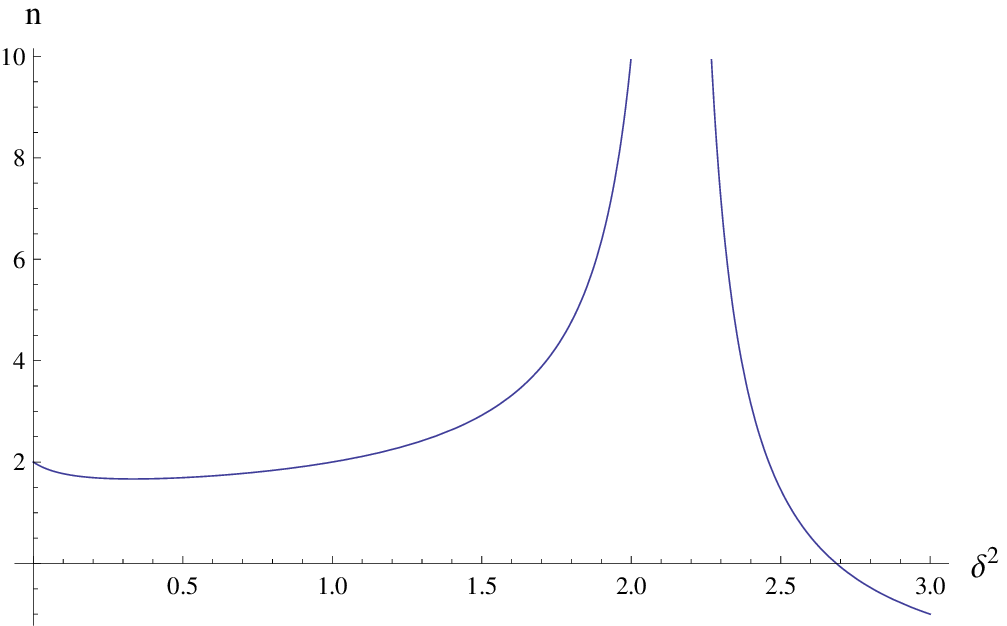}
\caption{The scaling exponent $n$ for $\gamma\delta  = 1 $ of the optical conductivity $\sigma (\omega) \sim \omega^n$ as a function of $\d^2\leq 3$.}
\label{gd1ACconductivity}
\end{minipage}
\hfil
\begin{minipage}{35mm}
\includegraphics[width=\linewidth,height=30mm]{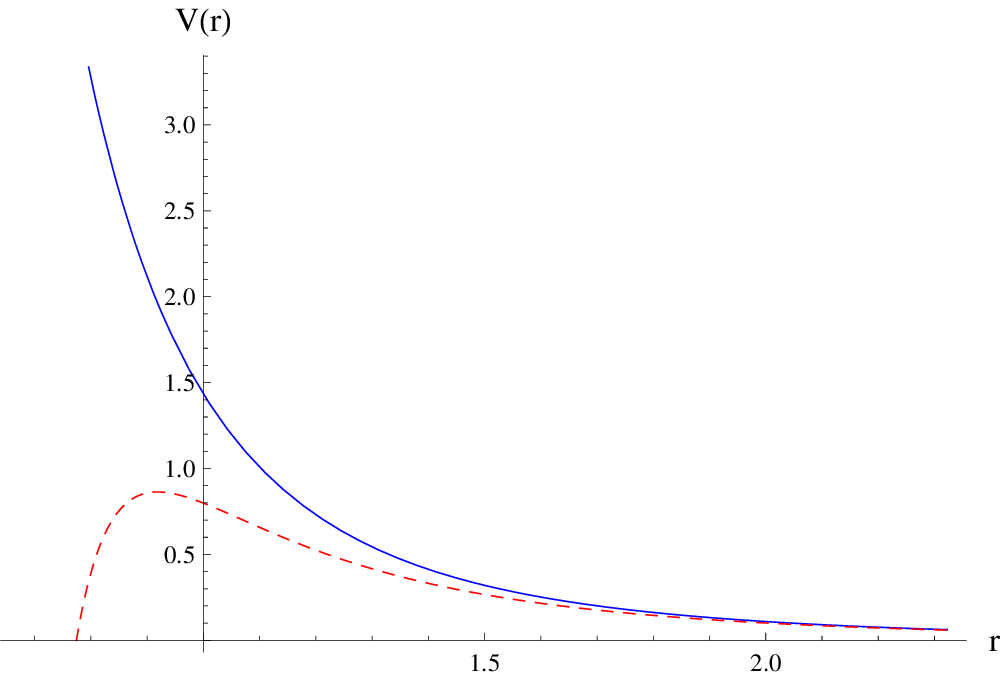}
\caption{Schr\"odinger potential for fluctuation at extremality (solid) 
and away from it (dashed).}
\label{Fig:gd1VSupper}
\end{minipage}
\hfil
\begin{minipage}{35mm}
\includegraphics[width=\linewidth,height=30mm]{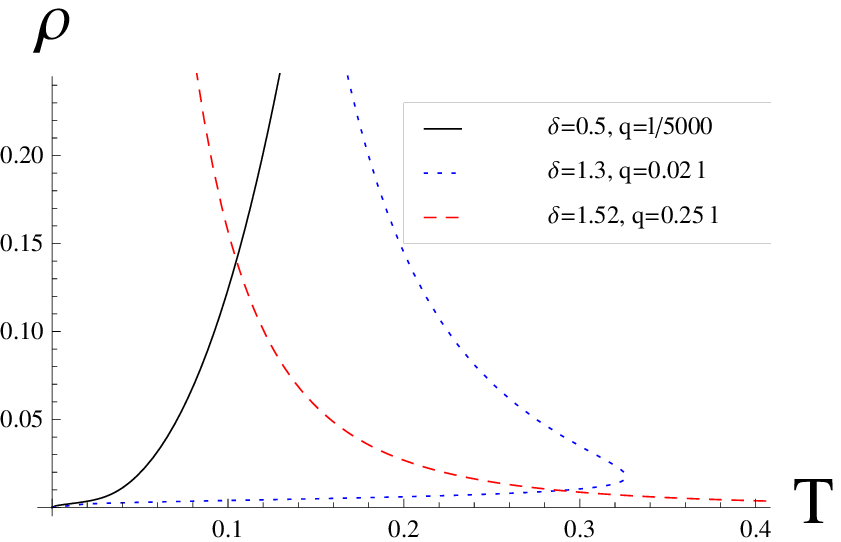}
\caption{DC resistivity for $\gamma\delta  = 1 $ with lower (solid), intermediate (dotted) and upper (dashed) ranges.}
\label{dg=1Resistivity}
\end{minipage}
\hfil
\begin{minipage}{35mm}
\includegraphics[width=\linewidth,height=30mm]{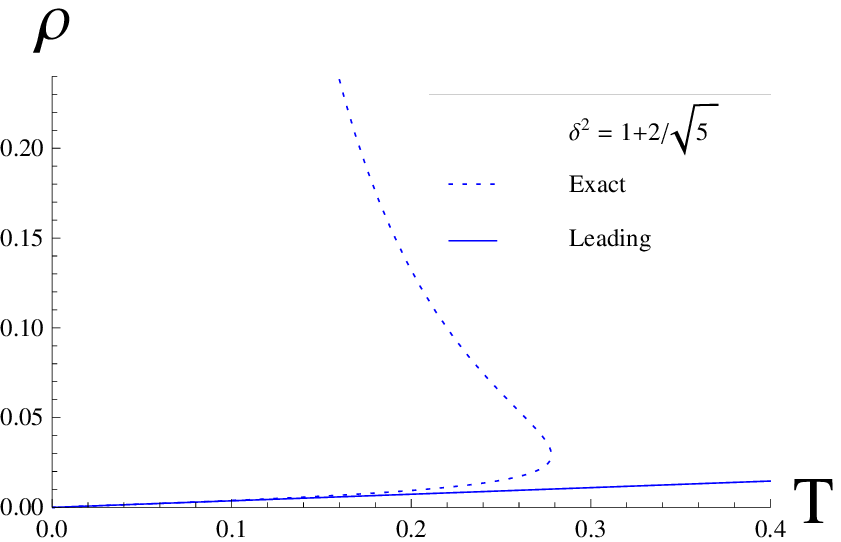}
\caption{DC resistivity for $\delta^2 = 1+ 2/\sqrt{5}$ gives the linear temperature dependence to leading order. }
\label{Fig:dg=1-d=1.3763-DC-Exact-Leading}
\end{minipage}
\end{figure}

We will in turn consider four different classes of solutions : uncharged, near extremal, exact solution with 
$\gamma \delta =1$ and with $\gamma =\delta$. 
The thermodynamic properties and phase structure of these solutions are summarized in \cite{Rene}. 
The exact solutions were found earlier in \cite{Charmousis:2009xr}.
\begin{itemize}
\item Uncharged solution : The solution depends on one parameter $\delta$ and it has an acceptable 
singularity when $0 \leq |\d| < 1$ according to the criteria in the previous section.  
We analyze the spectrum of current excitations with probe Maxwell term 
with a parameter $\gamma'$ on the uncharged background. When $\frac{\gamma}{\delta} > \frac{3}{2}$ or 
$\frac{\gamma}{\delta} < -\frac{1}{2}$ and the UV dimension of the scalar $\Delta < 1$, the
Schr\"odinger potential diverges both in the UV and the IR. Thus the spectrum is discrete
and gaped, which resembles to an insulator. Otherwise it is a conductor.
When $-\frac{1}{2} < \frac{\gamma}{\delta} < \frac{3}{2}$, the spectral problem is unacceptable and 
our EHT becomes unreliable. 

When $|\delta| <1$, AC conductivity is given by $\sigma \sim \omega^{\sqrt{4c+1} -1}$ with 
$c= \frac{(\g\d + 1-\d^2)\g\d}{(1-\d^2)^2}$. This reveals interesting scaling behavior 
$$ \sigma \sim \omega^{-\frac{2}{3}} \qquad \text{when} \qquad \gamma = \frac{\d^2 - 1}{3\d} 
\quad \text{or} \quad \frac{2(\d^2 - 1)}{3\d} \;,
$$
which seems to fit to an available experimental data \cite{nature}.
 
At the lowest densities the DC conductivity shows  
$
\rho=1/\sigma \sim T^{{2(p-1)\gamma\d+2(p-3)\over (p-1)\d^2-2}}\,.
$
In particular, in this dilute regime it is independent of the dilatonic nature of the scalar field, 
which can be parametrized by a parameter $k$. Thus the conductivity can be linear in temperature $T$ for
$$ 
\rho = \frac{1}{\s} \sim T \qquad \text{when} \qquad  \gamma= \frac\d2-\frac{1}{2\d }\,,
$$
which is universal for unconventional superconductor at the optimal doping \cite{Ong1991}. 
This DC resistivity has an interesting relation with entropy and heat capacity in temperature $T$ 
$
\rho~~\sim~~ S~~\sim~~ C_V ~\sim~~T\,.
$ 
As the zero charge density system never exhibits an entropy linear in temperature, 
we conclude that when linear resistivity appears, it is due to the presence of charge carriers.
For higher charge density, 
$
\rho \sim T^{2k\d+2\over 1-\d^2}\langle J^t\rangle\,,
$
where $\langle J^t\rangle$ is charge density. In this  case, the resistivity can never be linear.

\item Near extremal solution : We view this solution as near extremal limit of full general solutions of 
the action (\ref{ActionLiouville}) with two parameters $(\gamma, \delta)$ describing our 
effective holographic theory deep in IR. These are the near extremal approximations to the two exact 
solutions with $\g\d =1$ and $\g=\d$. It contains the Lifshitz solution as a special case with 
$\g = -\sqrt{4/(z-1)}$ and $\d=0$ \cite{Taylor:2008tg}. Entropy vanishes at extremality if $\g \neq \d$.

AC conductivity is given by $\sigma \sim \omega^n$ with $n = \left|\frac{-12+(\d-\gamma)(3\gamma 
+ 5\d)}{-4 + (\d - \gamma) (\gamma + 3 \d )}\right| - 1$. It turns out that 
$n$ is positive for the regions where the solutions is thermodynamically stable and completable to AdS space. 
If $3\ga^2-\da^2-2\ga\da+4 > 2(\gamma-\delta)^2$, which is equivalent to have vanishing Schr\"odinger 
potential in IR, the spectrum of charged fluctuation is gapless and continuous, describing a conductor.
If $3\ga^2-\da^2-2\ga\da+4 < 2(\gamma-\delta)^2$ and $n>0$ with the assumption to have AdS boundary and 
$1=2 < \Delta_\phi < 1$ there, scalar potential diverges both at UV and IR. This describe a finite charge density
system whose charge excitation spectrum has a gap and is discrete, which seems to describe a $(2+1-$dimensional 
Mott Insulator.

DC conductivity, on the other hand, $\rho \sim T^{m}$ with 
$m= {4k (\d-\gamma)+2(\d-\gamma)^2\over 4(1-\d(\d-\gamma))+(\d-\gamma)^2}$. 
Thus we have linear temperature dependence of resistivity for  
$\gamma_{\pm}=3\d+2k\pm 2\sqrt{1+(\d+k)^2}$. For $k=0$, temperature dependence of the entropy, 
heat capacity and the resistivity are the same, $\gamma_{\pm}$, $\rho(T) \sim S(T) \sim C_v(T)\sim T$, 
which is linear in $T$.

The leading scaling behavior of thermodynamic functions and conductivities is expected to be universal and be valid for all solutions for this class of theories near extremality. Therefore, they provide a profile that might help select EHTs as descriptions of concrete CM systems.

\item Exact solution with $\gamma \delta =1$ : For $0 \leq \d^2 < 1+ 2/\sqrt{3}$, 
the AC conductivity at extremality scales as
$
	\sigma(\omega) \simeq \omega^n \,,$ with $n =  \frac{(3-\d^2)(5\d^2+1)}{|3\d^4-6\d^2-1|} - 1\,.
$	
The exponent is always larger than $5/3$ in this region, and diverges at $\d^2 = 1+2/\sqrt{3}$. 
The spectrum is continuous and has no mass gap in this region and thus behaves as a conductor.
For $1+2/\sqrt{3}<\delta^2 < (5+\sqrt{33})/4$, the Schr\"odinger potential diverges 
near the horizon at extremality, and hence the system is insulating, while it becomes conducting away from 
extremality (see \Figref{Fig:gd1VSupper}).
For $(5+\sqrt{33})/4<\d^2<3$, the system is conducting with a continuum of states at extremality.

The DC conductivity for massive charge carriers is depicted in \Figref{dg=1Resistivity}. In the lower range $0\leq\delta^2\leq 1$, 
the DC resistivity vanishes at zero temperature and then rises in the black hole phase.
In the intermediate range $1<\delta^2 <1+{2\over \sqrt{3}}$, the resistivity drops steeply with temperature 
in the unstable large black hole branch and rises slowly with temperature in the thermodynamically dominant small black hole branch. 
In this latter branch, the low-temperature resistivity is linear for $\delta^2 = 1+ 2/\sqrt{5}$. 
The validity of linear regime is determined by the IR dynamical scale, $\ell$.  
Finally, in the upper branch $1+2/\sqrt{3}<\delta^2 <3$, the resistivity is decreasing with the temperature.
In the special case $\delta^2=1$, the resistivity is finite at zero temperature, then increases, and finally diverges at a higher temperature signaling potentially a critical behavior.

\item Exact solution with $\gamma =\delta$ : As explained in \cite{Rene}, the upper range of $\g\d=1$ case 
has no equivalent in $\gamma =\delta$. The zero temperature AC conductivity for all such black holes behaves 
as $\omega^2$ independent of $\delta$. In both ranges the resistivity is non-zero at zero temperature and has a regular 
low-temperature expansion in integer powers of the temperature starting with a linear term. Transport properties of 
$\g=\d$ case are similar to those of $\g\d=1$ case in an appropriate range.   
\end{itemize}

\vspace{-0.1in}
\section{Outlook}

After \cite{Charmousis:2010zz}, there are several attempts to investigate the transport properties of the dilatonic 
background. The near extremal solutions is also considered in \cite{Lee:2010ii}. 
For the important case with black branes carrying both electric and magnetic charges in
Einstein-Maxwell theory coupled to a dilaton-axion in asymptotically anti de Sitter space is analyzed in \cite{Goldstein:2010aw}.

It is worthwhile to mention recent attempts, using holographic finite density systems, to describe 
the benchmark conductivity features of unconventional superconductors: a linear temperature dependence of 
resistivity, $\rho \sim T$, at the optimal doping, its crossover to $T^2$ in the over-doped region, 
and a quadratic temperature dependence of inverse Hall angle, 
$\cot \Theta_H = \frac{\sigma^{xx}}{\sigma^{xy}} \sim T^2$. 
The linear temperature dependence of the Ohmic resistivity was found in spaces with AdS or Lifshitz scaling
\cite{Faulkner:2010zz, Hartnoll:2009ns, Charmousis:2010zz, Lee:2010ii},
as well as Schr\"odinger space with special conditions \cite{Ammon:2010eq,Kim:2010tf}.
A linear resistivity and a crossover to quadratic was found in a larger class of scaling geometries in
\cite{Charmousis:2010zz}. The Hall angle behavior was recently discussed
using Lifshitz type metric with broken rotational symmetry \cite{Pal:2010sx}. Finally 
these all three benchmark behaviors including a crossover were reproduced in a single holographic system 
in \cite{Kim201012}. 

We discuss a set of effective holographic theories parametrized by two exponents $(\gamma,\delta)$ associated with 
dilatonic coupling to the gauge fields and potential. The parameter space is severely constrained by the nature of 
the IR singularity. Imposing constraints for the singularity to be a good kind removes part of the $(\gamma,\delta)$ 
plane as discussed. The transport properties of the allowed parameter space are studied and characterized by their  
scaling behavior. They suggest that there are EHTs that display the hallmark behavior of general strange metals, 
namely linear resistivity. Moreover, the DC resistivity is correlated to the entropy and the specific heat.

There are however more features of the physics of the EHTs that need to be analyzed so that their suitability 
as theories of strange metal behavior can be assessed. An important general ingredient are the zero and low 
temperature spectra of fluctuations, that should be derived in order to completely characterize the low energy 
degrees of freedom, as well as the energy-energy and charge-charge correlators.
In particular the interplay between insulating versus conducting behavior must be further analyzed. 
It is a generic property of the holographic systems described here, in the $(2+1)-$dimensional case, 
to favor conducting behavior. Indeed, with the exception of strongly relevant dynamics, the systems 
at small charge density seem to be conductors at any small but non-zero temperature.
There is generically a continuous phase transition to the extremal solution which is insulating.

The phase structure also seems to be interestingly varying with the IR ``strength" of the scalar operator 
captured by the exponent $\delta$. There are indications from our analysis that the appearance of discrete 
spectra at low temperature is ``delayed" by the finite charge density.
At zero charge density, such spectra appear when $\d^2>1$ while at finite density in the $\gamma=\delta$ case 
they never appear, and for $\gamma\delta=1$ their appearance is delayed 
until $\delta^2=1+{2\over \sqrt{3}}$. This structure needs verification from a detailed analysis of low energy spectra.

BSK would like to thank to the organizers for a stimulating environment. 
We are grateful to E. Kiritsis and C. Panagopoulos for the discussions and for their advice.  
This work was partially supported by a European Union grant FP7-REGPOT-2008-1-CreteHEPCosmo-228644, 
Excellence grant MEXT-CT-2006-039047 and by ANR grant STR-COSMO, ANR-09-BLAN-0157.

\end{document}